\documentstyle[aps,twocolumn]{revtex}
\begin{document}
\draft
\title{Gravitational waves, black holes and cosmic strings 
in cylindrical symmetry}
\author{Sean A. Hayward}
\address{Center for Gravitational Physics and Geometry,
104 Davey Laboratory, The Pennsylvania State University,
University Park, PA 16802-6300, U.S.A.\\
{\tt hayward@gravity.phys.psu.edu}}
\date{Revised 6th February 2000}
\maketitle

\begin{abstract}
Gravitational waves in cylindrically symmetric Einstein gravity 
are described by an effective energy tensor 
with the same form as that of a massless Klein-Gordon field,
in terms of a gravitational potential generalizing the Newtonian potential.
Energy-momentum vectors for the gravitational waves and matter 
are defined with respect to a canonical flow of time.
The combined energy-momentum is covariantly conserved,
the corresponding charge being the modified Thorne energy.
Energy conservation is formulated as the first law 
expressing the gradient of the energy as work and energy-supply terms, 
including the energy flux of the gravitational waves.
Projecting this equation along a trapping horizon 
yields a first law of black-hole dynamics 
containing the expected term involving area and surface gravity,
where the dynamic surface gravity is defined 
with respect to the canonical flow of time.
A first law for dynamic cosmic strings also follows.
The Einstein equation is written as three wave equations plus the first law,
each with sources determined by the combined energy tensor 
of the matter and gravitational waves.
\end{abstract}
\pacs{04.30.-w, 04.70.-s, 04.20.Dw, 11.27.+d}

\section{Introduction}

The emission of gravitational waves by black holes 
is under intense investigation,
in preparation for the expected observational study of black holes 
and other astrophysical objects by upcoming gravitational-wave detectors.
Most of this work is either numerical or in an approximation,
since we lack a full physical understanding of the dynamics of black holes 
and gravitational waves.
The accepted theory of black holes mainly concerns statics and asymptotics;
for instance, there are standard asymptotic definitions of 
mass, angular momentum and the energy flux of gravitational waves,
with surface gravity defined for stationary black holes.
One might expect such physical quantities to play a more local role 
in dynamical processes.

A framework for addressing black-hole dynamics 
was introduced a few years ago\cite{bhd}.
Black holes were given a local, dynamical definition 
in terms of trapping horizons,
which are essentially the locations where light waves are marginally trapped
by the gravitational field.
Some basic properties of black holes were established,
such as that the horizon is achronal and therefore locally one-way traversible,
and a second law expressing the increase of area of the trapping horizon.
A comprehensive picture of the spherically symmetric case 
has since been developed,
involving local, dynamical definitions of physical quantities 
such as energy\cite{sph} and surface gravity\cite{1st},
related by local, dynamical equations including a first law.

This article introduces the corresponding physical, geometrical quantities 
and equations in cylindrical symmetry,
which has the additional complexity of gravitational waves.
In vacuo, these are known as Einstein-Rosen gravitational waves\cite{ER}.
Of course, cylindrical black holes are rather unphysical,
but their study is a precursor to that of axially symmetric black holes,
in particular their interaction with gravitational waves.
The main conceptual issue is whether 
the energy flux of gravitational waves admits a local, dynamical definition.
This turns out to be so, indeed quite natural, 
in that there is a covariant conservation law 
for the combined energy-momentum of the gravitational waves and matter.

Due to some peculiar properties of black holes in cylindrical symmetry,
naked singularities may also be a feature of gravitational collapse.
Consequently the framework is extended to include axial singularities,
particularly dynamic cosmic strings.

The article is organised as follows.
Section II reviews cylindrical symmetry in the sense of Melvin\cite{M1,M2}
and defines regular axes and axial singularities.
Section III defines black holes and related ideas of gravitational trapping.
Section IV reviews the modified Thorne energy\cite{T1},
which turns out to be the appropriate definition of gravitational energy.
Section V introduces a canonical flow of time 
and the corresponding energy-momentum vector of the matter.
Section VI introduces the energy-momentum vector of the gravitational waves,
states the conservation law, 
and introduces the gravitational potential 
and effective energy tensor of the gravitational waves.
Section VII states the first law in terms of the energy flux 
of the gravitational waves, plus energy-flux and work terms for the matter.
Section VIII defines the surface gravity.
Section IX collects basic laws of black-hole dynamics,
including the projection of the first law along a trapping horizon,
which involves area and surface gravity as expected.
Section X discusses some static examples 
including black holes and cosmic strings.
Section XI defines dynamic cosmic strings 
and discusses issues of cosmic censorship and predictability.
Section XII concludes.
An Appendix derives the Einstein equation 
and lists various expressions in standard coordinate systems.
Einstein gravity is assumed,
though the application of the Einstein equation is stated as appropriate.

\section{Regular and singular axes}

There is sometimes discussion about how to distinguish 
cylindrical, planar and toroidal symmetry.
Each may be defined locally by the existence of 
two commuting, spatial, Killing vectors,
such that the orthogonal space is integrable.
Then there exist coordinates $(\varphi,z)$ such that the Killing vectors are 
$(\xi_\varphi,\xi_z)=(\partial/\partial\varphi,\partial/\partial z)$. 
Here it will also be demanded that 
each of the Killing vectors be hypersurface orthogonal,
which in planar symmetry is often called the polarized case.
Then {\em central cylindrical symmetry} may be defined by 
the existence of an axis, defined shortly, 
with respect to exactly one of the Killing coordinates, 
the azimuthal angle $\varphi$, which will be identified at 0 and $2\pi$.
Central toroidal symmetry may be defined by 
the existence of axes with respect to both coordinates,
and planar symmetry by no such axes.
For {\em local} cylindrical, toroidal and planar symmetry,
one requires only the identifications, 
without prejudice as to the existence of axes.
This may be useful since, for instance, there may be black holes 
which look cylindrically symmetric from the outside, but have no axis inside.

The norms of the Killing vectors are geometrical invariants,
the {\em circumferential radius}
\begin{equation}
\rho=\sqrt{\xi_\varphi\cdot\xi_\varphi^\flat}
\end{equation}
and the {\em specific length}
\begin{equation}
\ell=\sqrt{\xi_z\cdot\xi_z^\flat}
\end{equation}
where the sign convention is that spatial metrics are positive definite,
the dot denotes contraction 
and the flat $\flat$ denotes the covariant dual 
with respect to the space-time metric, i.e.\ index lowering.
Similarly, a sharp $\sharp$ will denote the contravariant dual.
Here and throughout the article, 
specific refers to the need to scale quantities 
by some specific length of the infinite cylinder, 
which is arbitrary given the freedom to rescale $\xi_z$ by a constant factor.
Likewise, a cylinder will always mean a cylinder of symmetry.
Coordinates for the 2-dimensional quotient space, 
orthogonal to the cylinders, will not be taken in the main text,
in order to stress the geometrical invariance of the quantities and equations.
See the Appendix for a list of the required quantities 
expressed in a standard coordinate system.

An {\em axis} will be defined as a boundary of the quotient space-time 
coinciding with $\rho=0$, such that $\ell^{-1}=O(1)$.
Then $\ell$ is either finite and non-zero at the axis, or becomes infinite.
An axis will be said to be {\em regular} if $\ell$ is finite and
\begin{eqnarray}
&&\nabla\rho\cdot\nabla^\sharp\rho=1+O(\rho^2)\\
&&\nabla\rho\cdot\nabla^\sharp\ell=O(\rho)\\
&&\nabla\ell\cdot\nabla^\sharp\ell=O(1)
\end{eqnarray}
and {\em singular} otherwise, 
where $\nabla$ is the covariant derivative of the space-time metric.
A singular axis will also be called an {\em axial singularity}.
The first of the above conditions is standard, e.g.\cite{CSV}, 
but the others appear to be new.
Some such conditions are necessary, 
since the first condition alone allows curvature singularities at the axis, 
e.g.\ for $\ell=1+\rho+O(\rho^2)$.

This article will adopt 
the above definition of central cylindrical symmetry for definiteness, 
though local cylindrical symmetry suffices for some of the results.
This is more rigid than a recent suggestion of Carot et al.\cite{CSV}
which does not require the orthogonal space to be integrable.
One might call this twisted cylindrical symmetry.
This is analogous to the relaxation from static to stationary space-times 
and would be expected to allow angular momentum.
The present definition is essentially the whole-cylinder symmetry 
of Melvin\cite{M1,M2},
though it should be stressed that the axis has not been required to be regular.
In particular, cosmic strings are allowed.

\section{Black holes}

The function
\begin{equation}
r=\rho\ell
\end{equation}
plays a similar role to the areal radius of spherical symmetry, 
with $\ell$ scaling various physical quantities 
so that they have the physically correct units.
In particular,
\begin{equation}
A=2\pi r
\end{equation}
is the {\em specific area} of the cylinders and
\begin{equation}
V=\pi r^2
\end{equation}
is the corresponding volume.
The key ideas of gravitational trapping\cite{bhd} may then be defined 
directly in terms of $r$ or $A$, as in spherical symmetry\cite{1st}, as follows.
Here achronal will mean spatial or null
and causal will mean temporal or null.

A cylinder is said to be {\em trapped}, {\em marginal} or {\em untrapped} 
as $\nabla^\sharp r$ is temporal, null or spatial respectively.
Assuming a time orientation,
if $\nabla^\sharp r$ is future (respectively past) causal, 
then the cylinder is {\em future} (respectively {\em past}) trapped or marginal.
On an untrapped or marginal cylinder, 
an achronal normal direction is {\em outward} (respectively {\em inward})
if $r$ is increasing (respectively decreasing) in that direction.
This provides a local spatial orientation in an untrapped region.
A marginal cylinder is {\em outer}, {\em degenerate} or {\em inner}
as $\nabla^2r>0$, $\nabla^2r=0$ or $\nabla^2r<0$ respectively.
A {\em trapping horizon} is a hypersurface foliated by marginal cylinders.
A future (respectively past) outer trapping horizon is proposed 
as the local, dynamical definition of a black (respectively white) hole.
Trapping horizons define what some authors seem to call apparent horizons,
though the textbook definition of apparent horizon differs,
being the boundary of the region containing outer trapped surfaces
in an asymptotically flat spatial hypersurface\cite{HE,W}.

One may also say that an axial singularity is trapped (respectively untrapped)
if it has a neighbourhood of trapped (respectively untrapped) cylinders, 
cf.\ Chiba\cite{Ch}.
If the singularity is smooth (as a boundary in the quotient space) 
it can alternatively be defined to be trapped, marginal or untrapped 
as $\nabla^\sharp r$ is temporal, null or spatial respectively.
A local formulation of weak cosmic censorship would be that 
only trapped (or perhaps marginal) singularities can form 
from physically reasonable matter and initial data.
Strong cosmic censorship would require the formation of 
spatial (or future-null) singularities only,
where the causal nature of the singularity is defined by the quotient metric.
However, as in the spherically symmetric case\cite{sph}, 
it is straightforward to show that 
there is an equivalence between these strong and weak versions:
a smooth axial singularity is spatial, null or temporal 
as it is trapped, marginal or untrapped respectively.

\section{Energy}

In Thorne's paper on cylindrical energy\cite{T1}
there is a note added in proof to the effect that 
a certain modification of the definition renders it finite in space-time.
This modified definition, also used by Chiba\cite{Ch}, can be written as
\begin{equation}
E=(1-\ell^{-2}\nabla r\cdot\nabla^\sharp r)/8
\end{equation}
with units such that Newton's gravitational constant is unity.
It transpires that this is indeed the appropriate definition 
of {\em gravitational energy per specific length},
for reasons developed in the remainder of this article.
Principally, $E$ is the charge associated with a conserved current 
which combines the energy-momentum of the matter and gravitational waves.
For the moment, some other properties of $E$ will be noted.

The definition has a similar form to that of 
the energy in spherical symmetry, the Misner-Sharp energy\cite{sph}.
The two numerical coefficients are fixed so that 
$E$ vanishes in flat space-time 
and recovers the mass per specific length of a cylinder 
to leading order in the small-cylinder approximation at a regular axis,
as shown by Thorne\cite{T1} for the original definition,
the result for $E$ following directly.
In general, $E=O(\rho^2)$ at a regular axis, 
directly from the regularity conditions.
It also follows that $E$ has the correct Newtonian limit,
the mass per specific length of the cylinder.
An example of post-Newtonian behaviour is given later 
for the Levi-Civit\`a metric\cite{LC}.

A cylinder is trapped, marginal or untrapped
as $E>1/8$, $E=1/8$ or $E<1/8$ respectively.
This is similar to the condition in spherical symmetry\cite{sph},
but with a numerical value arising because 
$E$ has units of energy over length. 
The value 1/8 seems to be peculiar to cylindrical symmetry.
The corresponding term in $E$ may be regarded as a topological term, 
not present in the corresponding energy in planar symmetry,
which is essentially the Hawking energy\cite{H}.

\section{Time}

A {\em canonical flow of time} in cylindrical symmetry 
is generated by the vector
\begin{equation}
k=({*}dr)^\sharp
\end{equation}
where $d$ is the exterior derivative 
and $*$ the Hodge operator of the quotient space, 
i.e.\ ${*}d$ is a curl.
The sign of the curl can be fixed in an untrapped region 
so that $k$ is future-causal.
One may define $k$ equivalently, up to sign, by
\begin{eqnarray}
&&k\cdot\nabla r=0\\
&&k\cdot k^\flat=-\nabla r\cdot\nabla^\sharp r.
\end{eqnarray}
Then a cylinder is trapped, marginal or untrapped 
as $k$ is spatial, null or temporal respectively.
In particular, trapping horizons may equivalently be defined as 
hypersurfaces where $k$ is null.
In these and other ways,
$k$ is analogous to the Killing vector of a stationary space-time
or the Kodama vector of a spherically symmetric space-time\cite{1st}.

Since the divergence of a vector $v$ normal to the cylinders is given by
$\nabla\cdot v=r^{-1}{*}d{*}(rv^\flat)$,
it follows that $k$ is covariantly conserved:
\begin{equation}
\nabla\cdot k=0.
\end{equation}
This conserved current therefore admits a charge
\begin{equation}
Q[k]=-\int_\Sigma\star\cdot k
\end{equation}
which depends only on the boundary $\partial\Sigma$ 
of a spatial hypersurface $\Sigma$ with regular axis.
Here $\star$ is the space-time volume 4-form.
This charge is found to be the areal volume:
\begin{equation}
Q[k]=\int Vdz.
\end{equation}

The {\em energy-momentum density per specific length of the matter},
referred to the canonical flow of time, is given by the vector
\begin{equation}
j[T]=-\ell^{-2}(T\cdot k)^\sharp
\end{equation}
where $T$ is the energy tensor of the matter in covariant form.
In spherical symmetry, the analogous $j[T]$ is also conserved,
but this is generally not so in cylindrical symmetry.
The physical reason is that gravitational waves carry energy.
One might then wonder 
whether there is a definition of energy-momentum for the gravitational waves, 
to be added to $j[T]$ such that the combined energy-momentum is conserved.
It transpires that this is so, as shown in the next section.

\section{Energy-momentum of gravitational waves and the gravitational potential}

The {\em energy-momentum density per specific length} turns out to be
\begin{equation}
j=A^{-1}({*}dE)^\sharp.
\end{equation}
This is the modified version of the vector introduced by Thorne\cite{T1},
also used by Chiba\cite{Ch}.
Like $k$, it is covariantly conserved:
\begin{equation}
\nabla\cdot j=0.
\end{equation}
This conserved current therefore also admits a charge
\begin{equation}
Q[j]=-\int_\Sigma\star\cdot j
\end{equation}
which is found to be the energy:
\begin{equation}
Q[j]=\int Edz.
\end{equation}
Applying the Einstein equation, $j$ does indeed have a term in $j[T]$,
as one can see explicitly in standard coordinates in the Appendix.
Remarkably, the remaining term can be written in the same form.
That is, two components of the Einstein equation may be written as
\begin{equation}
j=j[T]+j[\Theta]
\end{equation}
where
\begin{equation}
j[\Theta]=-\ell^{-2}(\Theta\cdot k)^\sharp
\end{equation}
and
\begin{equation}
8\pi\Theta=2\nabla\phi\otimes\nabla\phi-(\nabla\phi\cdot\nabla^\sharp\phi)g
\end{equation}
where $\otimes$ denotes the symmetric tensor product,
$g$ is the space-time metric and
\begin{equation}
\phi=-\ln\ell.
\end{equation}
Then $j[\Theta]$ is the {\em energy-momentum density per specific length 
of the gravitational waves},
which when added to $j[T]$ yields the conserved current $j$.

One may therefore interpret $\Theta$ as 
the {\em effective energy tensor of the gravitational waves}.
Such effective energy tensors are known in the linearized approximation
and the short-wave approximation\cite{MTW}, but the above result is exact:
$\Theta$ is an invariantly defined tensor.
As shown in the Appendix,
the Einstein, Tolman and Landau-Lifshitz energy-momentum complexes
as calculated by Rosen and Virbhadra\cite{RV,V}
also partially agree with $\Theta$,
but are not invariant tensors.
One may recognise $\Theta$ as 
the energy tensor of a massless Klein-Gordon field $\phi$.
Thus $\phi$ plays the role of a potential for the gravitational waves.

Note that only the components of $\Theta$ orthogonal to the cylinders
contribute to $j[\Theta]$, but it transpires that the other components 
simplify the remaining components of the Einstein equation,
principally the 2-dimensional wave equation
\begin{equation}
{*}d{*}d\gamma=8\pi(T^\varphi_\varphi+\Theta^\varphi_\varphi)
\end{equation}
where $\gamma$ is defined by the determinant of the quotient metric:
\begin{equation}
e^{2\gamma}=\ell^2\sqrt{-\det}.
\end{equation}
Although $\gamma$ is not invariantly defined,
its 2-dimensional Laplacian ${*}d{*}d\gamma$ is invariant.
Since $T^\varphi_\varphi$ is the azimuthal pressure,
the above equation indicates that $\Theta^\varphi_\varphi$ 
acts as an azimuthal pressure due to the gravitational waves.

Another component of the Einstein equation may be written as
a wave equation for $\phi$ 
with another invariant of the energy tensor T as a source:
\begin{equation}
\nabla^2\phi=4\pi\varrho
\end{equation}
where
\begin{equation}
\varrho=-\hbox{tr}\,T-T^\varphi_\varphi+T^z_z
\end{equation}
and the trace refers to the quotient space.
This may also be written with $T+\Theta$ replacing $T$, since
\begin{eqnarray}
&&\hbox{tr}\,\Theta=0\\
&&\Theta^\varphi_\varphi=\Theta^z_z.
\end{eqnarray}
The same is true for the one component of the Einstein equation yet to be given.
Thus $\Theta$ is an effective energy tensor in a general sense:
writing the Einstein equation with $T+\Theta$ on one side,
instead of the usual $T$, simplifies the other side.

The wave equation for $\phi$ is recognisable as 
a relativistic version of the Poisson equation of Newtonian gravity,
with $\phi$ playing the role of the Newtonian gravitational potential.
This was pointed out by Melvin\cite{M1,M2},
though with the opposite sign for the potential.
The sign is determined by the fact that 
$\varrho$ reduces to the density in the Newtonian limit.
In summary, the {\em gravitational potential} $\phi$ both generalizes 
the Newtonian potential and acts as a potential for the gravitational waves.

\section{Unified first law}

The {\em first law} is the energy-balance equation 
expressing the gradient of $E$ according to the Einstein equation.
This can be written in a similar form to that of spherical symmetry,
\begin{equation}
dE=A\psi+\ell^{-2}wdV
\end{equation}
in terms of the {\em work density} (an energy density)
\begin{equation}
w=-\hbox{tr}\,T/2
\end{equation}
and the {\em energy flux per specific length}
\begin{equation}
\psi=\psi[T]+\psi[\Theta]
\end{equation}
which has been divided into contributions 
from the matter and the gravitational waves:
\begin{eqnarray}
&&\psi[T]=\ell^{-2}(T\cdot\nabla^\sharp r+w\nabla r)\\
&&\psi[\Theta]=\ell^{-2}\Theta\cdot\nabla^\sharp r.
\end{eqnarray}
These energy fluxes are essentially duals of the energy-momentum densities:
\begin{eqnarray}
&&j^\flat[T]={*}\psi[T]+\ell^{-2}wk^\flat\\
&&j^\flat[\Theta]={*}\psi[\Theta].
\end{eqnarray}
The terms in the first law involving $\psi$ and $w$ 
may be interpreted as energy-supply and work terms respectively,
analogous to the heat supply and work 
in the classical first law of thermodynamics\cite{th}.
This particular division into two terms is motivated by 
properties of $\psi$, described below, 
which make it analogous to heat flux in thermodynamics.
The unified first law was so called because it was shown in spherical symmetry 
that projecting it along the flow of a thermodynamic fluid 
yields a first law of relativistic thermodynamics,
while projecting it along a trapping horizon 
yields a first law of black-hole dynamics\cite{1st}.
In cylindrical symmetry, the above first law is also unified in the sense that 
it includes the energy flux of the gravitational waves.
It may also be applied to cosmic strings, as shown later.

The energy density and fluxes have similar properties
to those of the spherically symmetric case\cite{1st}.
Firstly, $\psi^\sharp[\Theta]$ is past (respectively future) causal
in future (respectively past) trapped regions,
and outward achronal in untrapped regions.
This last property of $\psi^\sharp[\Theta]$ is analogous to that of heat flux, 
which is a spatial vector in thermodynamics\cite{th}.
The same causal properties hold for $\psi^\sharp[T]$ 
assuming the null energy condition.
Then $\zeta\cdot\psi\ge0$ where $\zeta$ is an outward achronal vector.
Also $w\ge0$ assuming the dominant energy condition.
Then the first law implies
\begin{equation}
\zeta\cdot\nabla E\ge0.
\end{equation}
Thus $E$ is non-decreasing in any outward achronal direction 
in an untrapped region.
This {\em monotonicity} property immediately gives a {\em positivity} property:
on an untrapped achronal hypersurface with regular axis,
\begin{equation}
E\ge0.
\end{equation}
Similarly, on an untrapped achronal hypersurface with a cosmic string 
of energy per specific length $\varepsilon$, defined later,
\begin{equation}
E\ge\varepsilon.
\end{equation}
This type of argument can be used to give several inequalities 
for black holes\cite{in}.
If the space-time is asymptotically flat at spatial or null infinity,
one can define the asymptotic mass per specific length to be the limit of $E$, 
if it exists.
Then the same inequalities hold for the asymptotic mass.
The monotonicity property reduces at null infinity to a mass-loss property,
with $\psi$ reducing to a Bondi-like flux per specific length\cite{mon}.
Generally, $\psi$ is the outward flux minus the inward flux.

\section{Surface gravity}

A dynamic {\em surface gravity} $\kappa$ may be defined 
as in the spherically symmetric case\cite{1st} by
\begin{equation}
k\cdot(\nabla\wedge k^\flat)=\ell\kappa\nabla r
\end{equation}
where $\wedge$ denotes the antisymmetric tensor product.
That is, the 1-forms on each side of the equation are proportional
and $\kappa$ is defined as the proportionality constant.
Since $k=\pm\nabla^\sharp r$ on a trapping horizon,
this reduces on such a horizon to
\begin{equation}
k\cdot(\nabla\wedge k^\flat)=\pm\ell\kappa k^\flat
\end{equation}
which is analogous to the usual definition of stationary surface gravity,
with $k$ replacing the stationary Killing vector.
Then
\begin{equation}
2\ell\kappa={*}dk^\flat={*}d{*}d r.
\end{equation}
Since $\nabla^2r={*}d{*}dr$ on a marginal surface,
this shows that a trapping horizon is outer, degenerate or inner
as $\kappa>0$, $\kappa=0$ or $\kappa<0$ respectively.
This confirms a desired property of surface gravity,
that it should vanish for degenerate black holes.
As in spherical symmetry\cite{1st},
$\kappa$ is defined everywhere in the space-time, not just on horizons.

The remaining component of the Einstein equation 
can be written as a 2-dimensional wave equation for $r$:
\begin{equation}
{*}d{*}d r=-16\pi rw.
\end{equation}
Then
\begin{equation}
\kappa=-8\pi\rho w.
\end{equation}
Therefore black or white holes, as defined by outer trapping horizons,
are not consistent with the dominant energy condition.
This is implied by the general {\em topology law} for black holes\cite{bhd}
and reflects an unphysical aspect of cylindrical black holes.
However, note that it is possible to maintain the null energy condition
while violating the dominant energy condition.
The simplest example is a negative cosmological constant $\Lambda<0$: 
$8\pi T=-\Lambda g$.

\section{Laws of black-hole dynamics}

Projecting the first law along a trapping horizon yields
\begin{equation}
0=\ell E'={\kappa A'\over{8\pi}}+\ell^{-1}wV'
\end{equation}
where $f'=\xi\cdot\nabla f$ in terms of a vector $\xi$ tangent to the horizon;
the quickest way to derive this uses the previous displayed equation.
Corresponding equations in the spherically symmetric case\cite{1st}
and the general case\cite{MH} 
have been called the {\em first law of black-hole dynamics},
as they take the same form as the familar first law of black-hole statics,
which actually is more like a Gibbs equation 
if one takes the thermodynamic parallel seriously\cite{th}.
In any case, the surface gravity and area occur as expected.
By analogy to the original definition of entropy by Clausius\cite{ent},
this identifies $A$ as a specific entropy and $\kappa$ as a temperature,
up to factors.
The fact that the entropy arises from the energy-supply term
again indicates that $\psi$ is analogous to heat flux,
which is proportional to entropy flux in conventional thermodynamics,
effectively defining temperature\cite{th}.

The {\em second law of black-hole dynamics}\cite{bhd} reads
\begin{equation}
A'\ge0
\end{equation}
assuming the null energy condition,
where the outward orientation of the future outer trapping horizon is such that 
the outward null expansion vanishes.
A trivial {\em zeroth law} expresses the constancy of $\kappa$ over a cylinder.
Of course, $\kappa$ generally varies along a trapping horizon,
unlike the static case.
The {\em signature law} of black-hole dynamics\cite{bhd} states that
an outer trapping horizon is achronal, again assuming the null energy condition.
Without the null energy condition,
outer trapping horizons may also be used to define 
traversible wormholes\cite{wh}.
Thus one may also investigate the interaction of wormholes 
and gravitational waves.

As indicated above, 
black holes in cylindrical symmetry have some peculiar features.
Firstly, their existence requires violation of the dominant energy condition,
whereas their usual properties require the null energy condition.
In axial symmetry, the dominant energy condition 
therefore implies a limit on the formation of arbitrarily spindly black holes,
as also argued by Thorne\cite{MTW,T2}.
Secondly, a black hole has constant energy per specific length, $E=1/8$.
In particular, 
there is an energy gap between black holes and flat space-time.
Thus it is impossible to smoothly develop a trapped region 
around an initially regular axis, where $E=0$.
Unless $E$ is badly behaved, e.g. jumping or becoming infinite,
gravitational collapse to a black hole first requires
the initially regular axis to develop into an axial singularity with $0<E<1/8$.
This is essentially a cosmic string, as the following examples indicate. 

\section{Static examples}

Static metrics can be locally specified by $(\ell,E,F)$ 
as functions of $r$, 
where $F=-\xi_t\cdot\xi_t^\flat$ in terms of the static Killing vector $\xi_t$.
Then the line-element is
\begin{equation}
ds^2=\ell^2dz^2+\ell^{-2}r^2d\varphi^2+\ell^{-2}(1-8E)^{-1}dr^2-Fdt^2.
\end{equation}
A simple example is the metric of a static cosmic string, 
which may be written as
\begin{equation}
ds^2=dz^2+r^2d\varphi^2+(1-8\varepsilon)^{-1}(dr^2-dt^2)
\end{equation}
where the constant $\varepsilon$ is interpreted as 
the energy per specific length of the string, since $E=\varepsilon$.
The space-time is locally flat, 
but has a conical axial singularity unless $\varepsilon$ vanishes.
Normally the case $0<\varepsilon<1/8$ is considered,
so that there is a positive angular deficit 
$\theta=2\pi(1-(1-8\varepsilon)^{1/2})$.
For small $\varepsilon$, $\theta=8\pi\varepsilon+O(\varepsilon^2)$,
agreeing to lowest order with the usual expression 
for the mass or energy of a cosmic string.
One may also consider $\varepsilon<0$ and $\varepsilon>1/8$,
the latter being a trapped, spatial string.

Cosmic strings are weak axial singularities in that $E$ is finite.
An example of a strong axial singularity is provided by 
the Levi-Civit\`a metric\cite{LC}.
This is the relativistic analogue of 
the Newtonian gravitational field outside an infinite cylinder,
given by the Newtonian gravitational potential $2m\ln r$,
where $m$ is the mass per specific length.
In Einstein gravity, the unique static vacuum solution has two constants,
due to the additional possibility of a superposed cosmic string.
The solution may be written, cf.\ Thorne\cite{T1}, as
\begin{eqnarray}
&&ds^2=r^{-4m}dz^2+r^{4m+2}d\varphi^2\cr
&&\qquad+(1-8m-8\varepsilon)^{-1}r^{4m(1+2m)}(dr^2-dt^2)
\end{eqnarray}
where the constants have been arranged so that 
$m$ is the mass per specific length 
and $\varepsilon$ is the energy per specific length.
That is, if one expands
\begin{equation}
E=1/8-(1/8-m-\varepsilon)r^{-8m^2}
\end{equation}
by assuming such units in the Newtonian limit,
i.e.\ $(r,m,\varepsilon,E)\mapsto(r,c^{-2}m,c^{-4}\varepsilon,c^{-4}E)$
where $c$ is the speed of light, one finds
\begin{equation}
E=mc^2+\varepsilon+m^2\ln r+O(c^{-2}).
\end{equation}
The third term has the same form as the expression $m\phi/2$ 
for Newtonian gravitational self-energy per specific length,
where $\phi=2m\ln r$ as in the Newtonian case.
Thus $E$ correctly encodes the mass of the line source,
its gravitational self-energy and the energy of the superposed string,
which may also be regarded as analogous to internal energy.
This allows a curious quasi-Newtonian distinction between energy and mass,
only the latter producing a non-flat gravitational field.

Finally, a simple example illustrating strings and black holes 
is the static metric given by
\begin{eqnarray}
&&ds^2=dz^2+r^2d\varphi^2+(1-8\varepsilon+a^2r^2)^{-1}dr^2\cr
&&\qquad-(1-8\varepsilon+a^2r^2)dt^2.
\end{eqnarray}
This is the trivial lift of the spinless 3-dimensional BTZ black hole\cite{BTZ},
which is a solution for a negative cosmological constant $\Lambda=-a^2$,
reducing to the 3-dimensional anti de~Sitter solution if $\varepsilon=0$.
Additional pressure terms are required in the 4-dimensional case\cite{LZ}.
If $a=0$, a static cosmic string is recovered,
as above with a different scaling of $t$.
Generally
\begin{equation}
E=\varepsilon-a^2r^2/8
\end{equation}
so that there is a string of energy $\varepsilon$ per specific length,
with the other term in $E$ being the contribution of the cosmological constant,
analogously to the spherically symmetric case\cite{N}.
If $\varepsilon<1/8$, the string is naked.
If $\varepsilon>1/8$ there is a trapping horizon at 
$r=a^{-1}(8\varepsilon-1)^{1/2}$.
This is an outer trapping horizon with surface gravity $\kappa=a^2r$.
Thus there is a black hole 
with a similar global structure to the Schwarzschild black hole.
In particular, the string inside the black hole is spatial and trapped.

It was originally argued that 
the Euclidean version of the BTZ black hole yields 
a temperature $\kappa/2\pi$, as expected, and an entropy $4\pi r$\cite{BTZ}.
However, this involved a mass defined as $8\varepsilon-1$.
Repeating the argument with $\varepsilon$ as the mass 
yields an entropy $\pi r/2$,
or in the 4-dimensional case a specific entropy $A/4$. 
This expected value for entropy has since been confirmed 
by statistical-mechanical methods\cite{Cp}.

\section{Cosmic strings}

Since cosmic strings may be a feature 
in gravitational collapse in cylindrical symmetry,
it may be useful to consider dynamic strings.
Here a {\em cosmic string} will be defined as a smooth axial singularity 
for which $E$ has a finite limit, unique at each point
(in the boundary of the quotient space).
This limit 
\begin{equation}
\varepsilon=\lim_{r\to0}E 
\end{equation}
is the energy per specific length of the string.
Then a string is trapped and spatial if $\varepsilon>1/8$,
marginal and null if $\varepsilon=1/8$,
and untrapped and temporal if $\varepsilon<1/8$, cf.\ Chiba\cite{Ch}.
In particular, unlike the spherically symmetric case\cite{sph}
there is now a range of positive energy, $0<\varepsilon<1/8$,
for which axial singularities are locally naked.
Thus if gravitational collapse to a black hole is to occur,
then provided $\varepsilon$ is smooth, 
the initially regular axis must first develop into a cosmic string 
with small energy $\varepsilon$.
Then only if $\varepsilon$ increases past 1/8 will a trapped region form,
hiding the string.
Until or unless that happens, the string is a locally naked singularity.

This reflects the ease with which naked singularities seem to form 
in cylindrical symmetry\cite{Ch,T2,E} or axial symmetry\cite{MTW,ST1,ST2}, 
at least for dust.
This energy gap between flat space-time and black holes can be seen as 
the physical basis for the apparent failure of cosmic censorship
in cylindrical symmetry.
However, this could also be ascribed to the pathological nature of dust,
since similar studies for a fluid with pressure indicate that 
a bounce occurs instead\cite{T2,P}.
Another possible resolution suggested by Nakamura et al.\cite{NSN} is that 
the energy of the incipient singularity is dispersed by gravitational waves,
presumably preventing its formation or minimizing its effect.
Studies of gravitational-wave emission prior to singularity formation 
have yielded quite different results: 
negligible emission\cite{Ch,ST1,ST2,P} or a strong burst\cite{E}.

These issues can be studied in the context of cosmic strings.
The first law projected along a string reads
\begin{equation}
\dot\varepsilon=\lim_{r\to0}Ak\cdot\psi
\end{equation}
where $\dot f=k\cdot\nabla f$.
This is the {\em first law for cosmic strings},
relating the increase or decrease of $\varepsilon$ 
with the absorption or emission, 
respectively, of gravitational waves or matter.
Clearly $\varepsilon$ is constant 
if the flux $k\cdot\psi$ is finite at the string,
as occurs for Einstein-Rosen waves and a massless Klein-Gordon field.
Such radiation seems not to feel the presence of the string at all,
the incoming waves passing through it and re-emerging,
with no gravitational collapse.
On the other hand, if gravitational collapse to an axial singularity does occur,
this is a cosmic string with time-dependent energy 
if the flux $k\cdot\psi$ is $O(r^{-1})$,
i.e.\ for finite energy supply $Ak\cdot\psi$.
A stronger axial singularity could conceivably occur, 
with effectively infinite energy.
Either way, predictability is lost 
unless there is a prescription for the emitted radiation.

One may take the view that naked singularities are not problematic 
if predictability can be maintained.
For instance, Clarke\cite{Cl} has defined generalized hyperbolicity 
in terms of existence and uniqueness of solutions to the wave equation,
and Vickers \& Wilson\cite{VW} have shown that 
a static cosmic string is hyperbolic in such a sense.
To handle a dynamic cosmic string, however,
it seems necessary to impose a boundary condition at the string.

Without an internal model for the string,
the simplest prescription is the {\em reflecting} boundary condition,
\begin{equation}
\lim_{r\to0}Ak\cdot\psi[\dots]=0
\end{equation}
applied separately for $\Theta$ and any types of matter contributing to $T$.
(If only the combined flux $\psi$ were so constrained,
energy transfer between gravitational waves and matter would be allowed
at the string,
again removing predictability).
This boundary condition means that 
any radiation striking the string is immediately reflected,
leaving the energy of the string constant:
\begin{equation}
\dot\varepsilon=0.
\end{equation}
This is automatically obeyed for Einstein-Rosen waves,
but would be inconsistent with gravitational collapse as outlined above;
some absorption of radiation by the string 
is necessary for its energy to increase from the zero of initial regularity.
An alternative boundary condition is the {\em absorbing} condition
\begin{equation}
\lim_{r\to0}Ak^-\psi_-[\dots]=0
\end{equation}
where $\partial_-$ is the future-inward null direction, as in the Appendix.
In this case 
\begin{equation}
\dot\varepsilon\ge0
\end{equation}
assuming the null energy condition,
with the string absorbing all matter and radiation which strike it 
and emitting nothing.
One might think of such irreversibility as a second law for cosmic strings.

Another possibility is a linear combination of the above conditions:
\begin{equation}
\lim_{r\to0}A(\epsilon k^+\psi_++k^-\psi_-)[\dots]=0
\end{equation}
where $\partial_+$ is the future-outward null direction 
and the reflection coefficient $\epsilon$ is specified between 0 and 1, 
for no reflection and total reflection respectively.
There are yet more possibilities, for instance in a dust collapse 
one might impose an absorbing condition for the dust,
but a reflecting condition for the gravitational waves.
The question of whether 
such naked singularities are strong sources of gravitational waves 
is presumably dependent on the choice of boundary condition.

\section{Conclusion}

In cylindrically symmetric Einstein gravity, 
the physical quantities may be summarized as follows,
omitting the recurring qualification per specific length.
The kinematical or gravitational quantities are 
the specific area $A$ (or $r=A/2\pi$),
the canonical time vector $k$, the gravitational potential $\phi$ 
(or the specific length $\ell=e^{-\phi}$,
or the circumferential radius $\rho=r/\ell$),
the energy $E$ and the surface gravity $\kappa$.
The effective energy tensor $\Theta$ is defined in terms of $\phi$,
leading also to the energy-momentum $j[\Theta]$ 
(or the energy flux $\psi[\Theta]$) of the gravitational waves.
The matter quantities are the pressures $T^z_z$ and $T^\varphi_\varphi$,
the energy density $w$ (or $\varrho$) and the energy-momentum $j[T]$
(or the energy flux $\psi[T]$),
all of which are invariants of the energy tensor $T$.

The five independent components of the Einstein equation may be written as 
the first law plus three wave equations,
each with sources given by invariants of $T+\Theta$.
They are collected here as
\begin{eqnarray}
&&dE=A\psi+\ell^{-2}wdV\\
&&{*}d{*}d r=-16\pi rw\\
&&{*}d{*}d\gamma=8\pi(T^\varphi_\varphi+\Theta^\varphi_\varphi)\\
&&\nabla^2\phi=4\pi\varrho.
\end{eqnarray}
Instead of the first law, one may use the equation expressing $j$, 
the energy-momentum associated with $E$, as $j[T+\Theta]$. 
Energy conservation is expressed by the covariant conservation law 
$\nabla\cdot j=0$.
In summary, the Einstein equation takes an invariant form, 
with the combined energy tensor $T+\Theta$ as a source.
In this sense, 
$\Theta$ is an effective energy tensor for the gravitational waves.

At a time when detection of gravitational waves is expected to be imminent,
it may seem odd to recall that there was a historical debate about whether 
gravitational waves carry energy, or are well defined at all in Einstein theory.
Einstein-Rosen waves played a significant role in this debate\cite{RV,V,R,WW}.
There were several attempts to construct pseudotensors for the energy,
which are usually rejected nowadays due to their non-invariant nature.
The modern viewpoint seems to be 
that expressed by Misner, Thorne \& Wheeler\cite{MTW},
that gravitational waves and their energy may be defined 
only in an approximate or averaged sense.
Thus it may come as a surprise that a genuine tensor $\Theta$ exists
which encodes the energy of the gravitational waves in cylindrical symmetry.
One may partly ascribe this to the symmetry, 
since averaging is trivial over the cylinders.
However, there has been no need to average over the remaining dimensions,
or to assume small amplitude or wavelength,
as can be seen explicitly for Einstein-Rosen waves in the Appendix.
The discovery of $\Theta$ should at least resolve the debate 
about the energy of Einstein-Rosen waves.

Basic laws of black-hole dynamics have been given,
including a first law with the expected term involving area and surface gravity.
There is already evidence that these determine entropy and temperature 
for static black holes\cite{Cp},
and it may be conjectured that this generalizes to dynamic black holes.
Recently there has been considerable interest in 
another generalization of static black holes, isolated horizons\cite{ABF},
which are a certain type of null trapping horizon.
In particular, a quantum-geometrical derivation of black-hole entropy 
can be given for such local horizons\cite{ABK}.

Finally, there is a formula for the energy 
which unifies cylindrical and spherical symmetry:
\begin{equation}
M={C\over{4\pi}}\left(1-{\nabla A\cdot\nabla^\sharp A\over{16C^2}}\right).
\end{equation}
In spherical symmetry $M$ is the Misner-Sharp energy\cite{sph}
if $A$ is the area and $C$ the circumference of the spheres.
In cylindrical symmetry $M$ is the specific energy $\ell E$ if $C=\pi\ell/2$.
In each case, a sphere or cylinder is trapped if and only if
\begin{equation}
C<4\pi M.
\end{equation}
This is exactly the inequality in the hoop conjecture of Thorne\cite{T2}.
It is to be hoped that the theory developed here 
can be generalized to axial symmetry,
so as to have a fully dynamical description of rotating black holes 
and co-rotating black-hole collisions.

\bigskip\noindent
The author is grateful to Abhay Ashtekar 
and the Center for Gravitational Physics and Geometry for hospitality.

\section{Appendix}

In cylindrical symmetry as defined here, 
the line-element locally takes the form
\begin{equation}
ds^2=\ell^2dz^2+\rho^2d\varphi^2+h_{ij}dx^idx^j
\end{equation}
where $x^i$ are coordinates orthogonal to the cylinders 
and the metric functions $(\rho,\ell,h)$ are independent of $(\varphi,z)$.
Using double-null coordinates $x^\pm$ and definitions in the text, 
this takes the form
\begin{equation}
ds^2=e^{-2\phi}dz^2+e^{2\phi}r^2d\varphi^2-2e^{2\phi+2\gamma}dx^+dx^-.
\end{equation}
The remaining coordinate freedom consists of 
the null diffeomorphisms $x^\pm\mapsto\xi^\pm(x^\pm)$ 
and constant linear transformations $z\mapsto Bz+C$,
with $\varphi$ identified at 0 and $2\pi$.

Then for a function $f$ and a 1-form $\alpha$,
\begin{eqnarray}
&&(df)_\pm=\partial_\pm f\\
&&({*}\alpha)_\pm=\mp\alpha_\pm\\
&&{*}d\alpha=e^{-2\gamma-2\phi}(\partial_-\alpha_+-\partial_+\alpha_-)\\
&&\nabla\wedge\alpha=d\alpha/2\\
&&\nabla^2f={*}d{*}df+r^{-1}\nabla^\sharp r\cdot\nabla f\\
&&\qquad=-e^{-2\gamma-2\phi}\left(2\partial_+\partial_-f
+{\partial_+r\partial_-f+\partial_-r\partial_+f\over{r}}\right)
\end{eqnarray}
where the orientation of $*$ has been locally fixed so that
$\partial_+$ is future-outward and $\partial_-$ is future-inward 
on untrapped or marginal cylinders.
Then various quantities defined in the text may be written as
\begin{eqnarray}
&&E=(1+2e^{-2\gamma}\partial_+r\partial_-r)/8\\
&&k^\pm=\mp e^{-2\phi-2\gamma}\partial_\mp r\\
&&j^\pm=\mp A^{-1}e^{-2\phi-2\gamma}\partial_\mp E\\
&&j^\pm[T]=\pm e^{-2\phi-4\gamma}(T_{\mp\mp}\partial_\pm r-T_{+-}\partial_\mp r)\\
&&j^\pm[\Theta]=\pm e^{-2\phi-4\gamma}(\partial_\mp\phi)^2\partial_\pm r/4\pi\\
&&\Theta_{\pm\pm}=(\partial_\pm\phi)^2/4\pi\\
&&\Theta_{+-}=0\\
&&\Theta^\varphi_\varphi=\Theta^z_z
=e^{-2\phi-2\gamma}\partial_+\phi\partial_-\phi/4\pi\\
&&\star=e^{2\phi+2\gamma}rdz\wedge d\varphi\wedge dx^+\wedge dx^-\\
&&w=e^{-2\phi-2\gamma}T_{+-}\\
&&\psi_\pm[T]=-e^{-2\gamma}T_{\pm\pm}\partial_\mp r\\
&&\psi_\pm[\Theta]=-e^{-2\gamma}(\partial_\pm\phi)^2\partial_\mp r/4\pi\\
&&\kappa=-e^{-2\gamma-\phi}\partial_+\partial_-r.
\end{eqnarray}
The non-zero Christoffel symbols are
\begin{eqnarray}
&&\Gamma^\pm_{zz}=-e^{-4\phi-2\gamma}\partial_\mp\phi\\
&&\Gamma^\pm_{\varphi\varphi}=e^{-2\gamma}r(\partial_\mp r+r\partial_\mp\phi)\\
&&\Gamma^\pm_{\pm\pm}=2(\partial_\pm\phi+\partial_\pm\gamma)\\
&&\Gamma^\pm_{\mp\mp}=2(\partial_\mp\phi+\partial_\mp\gamma)\\
&&\Gamma^z_{z\pm}=-\partial_\pm\phi\\
&&\Gamma^\varphi_{\varphi\pm}=r^{-1}\partial_\pm r+\partial_\pm\phi
\end{eqnarray}
and the non-zero components of the Ricci tensor are
\begin{eqnarray}
&&R_{zz}=-e^{-4\phi-2\gamma}\times\cr
&&\qquad\left(2\partial_+\partial_-\phi
+r^{-1}(\partial_+r\partial_-\phi+\partial_-r\partial_+\phi)\right)\\
&&R_{\varphi\varphi}=e^{-2\gamma}r\times\cr
&&\qquad(2\partial_+\partial_-r+2r\partial_+\partial_-\phi
+\partial_+r\partial_-\phi+\partial_-r\partial_+\phi)\\
&&R_{\pm\pm}=
r^{-1}(2\partial_\pm r\partial_\pm\gamma-\partial_\pm\partial_\pm r)
-2(\partial_\pm\phi)^2\\
&&R_{+-}=-2\partial_+\partial_-\phi-2\partial_+\partial_-\gamma
-2\partial_+\phi\partial_-\phi\cr
&&\qquad\qquad-r^{-1}(\partial_+\partial_-r
+\partial_+r\partial_-\phi+\partial_-r\partial_+\phi).
\end{eqnarray}
The Einstein equation may then be written as
\begin{eqnarray}
&&\partial_\pm\partial_\pm r-2\partial_\pm r\partial_\pm\gamma
+2r(\partial_\pm\phi)^2=-8\pi rT_{\pm\pm}\\
&&\partial_+\partial_-r=8\pi rT_{+-}\\
&&\partial_+\partial_-\gamma+\partial_+\phi\partial_-\phi
=-4\pi e^{2\phi+2\gamma}T^\varphi_\varphi\\
&&2\partial_+\partial_-\phi
+r^{-1}(\partial_+r\partial_-\phi+\partial_-r\partial_+\phi)\cr
&&\qquad=4\pi\left(e^{2\phi+2\gamma}(T^\varphi_\varphi-T^z_z)-2T_{+-}\right).
\end{eqnarray}
It follows that
\begin{equation}
\partial_\pm E=2\pi re^{-2\gamma}\left(T_{+-}\partial_\pm r
-\left(T_{\pm\pm}+{(\partial_\pm\phi)^2\over{4\pi}}\right)\partial_\mp r\right).
\end{equation}
The null energy condition requires $T_{\pm\pm}\ge0$
and the dominant energy condition requires $T_{+-}\ge0$.
The equations in the main text follow straightforwardly 
from the above expressions.

Writing 
\begin{equation}
\sqrt{2}x^\pm=x^0\pm x^1
\end{equation}
puts the metric in the form of Melvin\cite{M2} and Thorne\cite{T1}.
The Einstein equation in these coordinates is found by linear combinations to be
\begin{eqnarray}
&&\ddot r-r''=-8\pi e^{2\gamma+2\phi}r(T^0_0+T^1_1)\\
&&\ddot\gamma-\gamma''=
\phi'^2-\dot\phi^2-8\pi e^{2\gamma+2\phi}T^\varphi_\varphi\\
&&\ddot\phi-\phi''+r^{-1}(\dot r\dot\phi-r'\phi')=
4\pi e^{2\gamma+2\phi}\times\cr
&&\qquad\qquad\qquad\qquad\qquad\qquad(T^0_0+T^1_1+T^\varphi_\varphi-T^z_z)\\
&&(r'^2-\dot r^2)\gamma'=-8\pi e^{2\gamma+2\phi}r(r'T^0_0+\dot rT^1_0)\cr
&&\qquad\qquad\qquad+rr'(\dot\phi^2+\phi'^2)
-2r\dot r\dot\phi\phi'+r'r''-\dot r\dot r'\\
&&(r'^2-\dot r^2)\dot\gamma=8\pi e^{2\gamma+2\phi}r(r'T^1_0-\dot rT^1_1)\cr
&&\qquad\qquad\qquad-r\dot r(\dot\phi^2+\phi'^2)
+2rr'\dot\phi\phi'-\dot r\ddot r+r'\dot r'
\end{eqnarray}
where $\dot f=\partial f/\partial x^0$ and $f'=\partial f/\partial x^1$
in this appendix only.
This is essentially the form given by Melvin\cite{M2} and Thorne\cite{T1},
except for misprints in their versions of the last equation.

If $w=0$, e.g.\ in vacuo, one may solve the wave equation for $r$ 
by taking $x^1=r$ locally in an untrapped region,
in which case the metric takes the Einstein-Rosen form\cite{ER}.
Note however that what most authors call $\psi$ has been replaced with $-\phi$,
since $\phi$ generalizes the Newtonian gravitational potential.
The energy tensor of the gravitational waves takes the form
\begin{eqnarray}
&&8\pi\Theta=(\dot\phi^2+\phi'^2)(dt\otimes dt+dr\otimes dr)
+2\dot\phi\phi'dt\otimes dr\cr
&&\qquad+(\dot\phi^2-\phi'^2)e^{-2\gamma}
(e^{-4\phi}dz\otimes dz+r^2d\varphi\otimes d\varphi)
\end{eqnarray}
where $t=x^0$.
The energy density $\Theta_{00}$ and energy flux $\Theta_{01}$ 
agree up to factors with 
the Einstein, Tolman and Landau-Lifshitz energy-momentum complexes
as calculated by Rosen and Virbhadra\cite{RV,V}.
The pressure terms seem to be new.

Einstein-Rosen waves are given by the Fourier-Bessel modes
\begin{equation}
\phi=\Re\int_0^\infty A(\omega)e^{i\omega t}J_0(\omega r)d\omega
\end{equation}
where $J_0$ is the zeroth-order Bessel function
and the function $A$ is complex.
This solves the wave equation for $\phi$ in vacuo,
the complete solution to the vacuum Einstein equation 
following by quadrature for $\gamma$.
For a single mode, substitution shows that $\Theta$ depends on 
the squares of the amplitude $|A|$ and frequency $\omega$.
In particular, the energy density $\Theta_{00}$ at the axis 
oscillates between 0 and $|A|^2\omega^2/8\pi$,
as for an electromagnetic wave.
Unlike the linearized or shortwave approximations\cite{MTW},
$|A|$ and $\omega$ can be arbitrarily large.
These fully non-linear waves still admit an energy tensor $\Theta$.

\end{document}